\documentclass[preprint,showpacs,preprintnumbers,amsmath,amssymb]{revtex4}
\usepackage{graphicx}
\usepackage{dcolumn}
\usepackage{bm}

\def\epjb{Eur. Phys. J. B }

\def\jap{J. Appl. Phys. }

\def\jvst{J. Vac. Sci. Technol. }

\def\sci{Science }

\newcommand{\afme}{{atomic force microscope}}
\newcommand{\df}{{depolarization field}}
\newcommand{\etal}{{\em et al}}
\newcommand{\LSMO}{{La$_{0.67}$Sr$_{0.33}$MnO$_3$}}
\newcommand{\PTO}{{PbTiO$_3$}}
\newcommand{\SRO}{{SrRuO$_3$}}
\newcommand{\STO}{{SrTiO$_3$}}

\begin{document}

\title{Monodomain to polydomain transition in ferroelectric \PTO\ thin films with \LSMO\ electrodes}

\author{C\'eline Lichtensteiger}
      \email{Celine.Lichtensteiger@physics.unige.ch}
\author{Matthew Dawber}
\author{Nicolas Stucki}
\author{Jean-Marc Triscone}
      \affiliation{DPMC - Universit\'e de Gen\`eve, 24 Quai Ernest-Ansermet, CH-1211 Gen\`eve 4, Switzerland}
\author{Jason~Hoffman}
\author{Jeng-Bang~Yau}
\author{Charles H. Ahn}
\affiliation{Department of Applied Physics, Yale University, New Haven, CT 06520-8284, USA}
\author{Laurent Despont}
\author{Philipp Aebi}
\affiliation{Institut de Physique, Universit\'{e} de Neuch\^{a}tel, CH-2000 Neuch\^{a}tel, Switzerland}
\date{\today}

\begin{abstract}

Finite size effects in ferroelectric thin films have been probed in
a series of epitaxial perovskite c-axis oriented PbTiO$_3$ films
grown on thin \LSMO\ epitaxial electrodes. The film thickness ranges
from 480 down to 28 \AA\ (7 unit cells). The evolution of the film
tetragonality $c/a$, studied using high resolution x-ray diffraction
measurements, shows first a decrease of $c/a$ with decreasing film
thickness followed by a recovery of $c/a$ at small thicknesses. This
recovery is accompanied by a change from a monodomain to a
polydomain configuration of the polarization, as directly
demonstrated by piezoresponse atomic force microscopy measurements.

\end{abstract}


\maketitle

Recently, both experimental and theoretical studies have suggested
that the critical size at which ferroelectricity disappears,
traditionally thought to be quite large, may actually be very
small~\cite{bun98,tyb99,gho00,mey01,str02,jun03_1,fon04,lic05,des06_1}. The depolarization field, which results from the imperfect
screening of the polarization, has been shown theoretically to play
a critical role~\cite{meh73,bat73_3,jun03_1}. In uniformly polarized
(monodomain) thin \PTO\ epitaxial films prepared on Nb-doped \STO\
substrates, it was experimentally shown that the increase of the
\df\ as the film thickness decreases leads to a reduction of the
polarization accompanied by a continuous reduction of the film
tetragonality $c/a$~\cite{lic05}. The direct relation between
tetragonality and polarization was also recently experimentally
demonstrated in \PTO /\STO\ superlattices~\cite{daw06}.

In this letter, epitaxial c-axis \PTO\ thin films with thicknesses
ranging from 480 \AA\ down to 28 \AA\ were grown on epitaxial \LSMO\
electrodes (typically 200-300 {\AA} thick) deposited onto (001)
insulating \STO\ substrates. It is found that the behavior of the
tetragonality is dramatically different from what is observed for
\PTO\ thin films prepared on metallic Nb-doped \STO\ substrates,
with an increase of $c/a$ observed for the thinnest film studied. We
show that this behavior is related to a change in the ferroelectric
domain structure, with the appearance of domains with 180$^\circ$
alternating polarization (polydomain configuration).

Using off-axis magnetron sputtering, extremely smooth (RMS surface
roughness of less than 2 \AA\ over 10$\times$10 $\mu$m$^2$ areas) \LSMO\
epitaxial thin films were grown
 on \STO\ substrates. Their ferromagnetic
$T_C$  is around or above 300 K and their resistivity at room
temperature typically 450 $\mu\Omega\cdot$cm. c-axis \PTO\ epitaxial
thin films of different thicknesses were then deposited on top of
the \LSMO\ electrodes~\cite{lic05}.

\begin{figure}[!ht]
 \begin{center}
  \begin{minipage}{6cm}
   \includegraphics[width=4cm,angle=90,keepaspectratio]{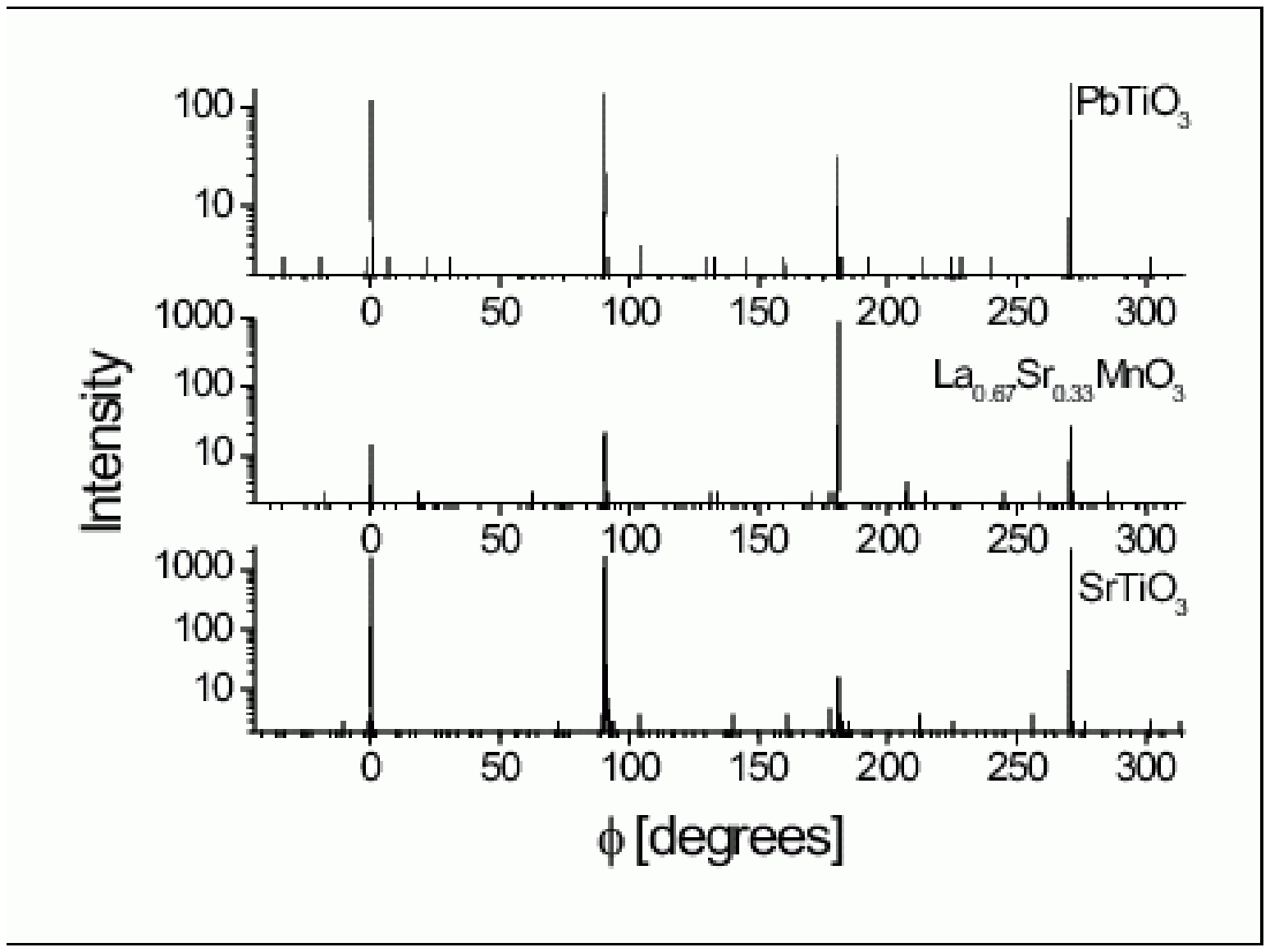}
  \end{minipage}
  \begin{minipage}{6cm}
   \includegraphics[width=4cm,angle=90,keepaspectratio]{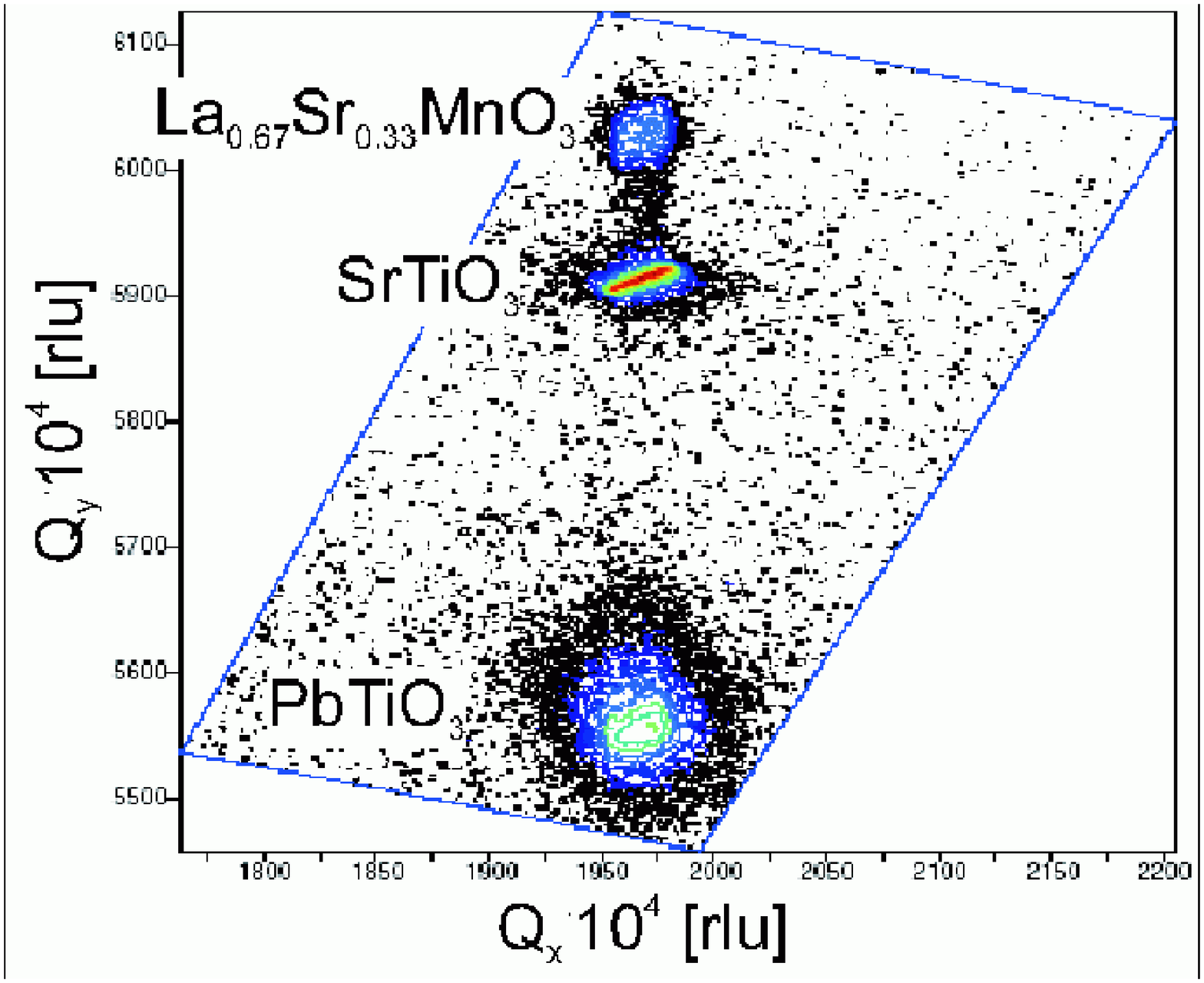}
  \end{minipage}
  \caption{\label{f:PoleFigure_qmap} X-ray diffraction
measurements on a 480 \AA\ \PTO / 220 \AA\ \LSMO\ bilayer. {\bf Left
-} $\phi$-scans demonstrating the ``cube on cube'' growth of \LSMO\
and \PTO\ on \STO .
  {\bf Right -} q-space map corresponding to scans in the (103) reciprocal-space region. The vertical alignment of the three peaks demonstrates that the
    whole structure is coherent.}
 \end{center}
\end{figure}

X-ray diffraction measurements were performed on these samples to
determine their thickness and lattice parameters.
Fig.~\ref{f:PoleFigure_qmap} (left) shows three $\phi$-scans around
the (101) family of planes obtained on a 480 \AA\ \PTO / 220 \AA\
\LSMO\ bilayer prepared on a \STO\ substrate, demonstrating the
tetragonal symmetry of the different materials and the ``cube on
cube'' growth of \PTO\ and \LSMO\ on the substrate and on top of
each other.

Coherent growth was demonstrated by q-space maps around the (103)
reflection, as shown in Fig.~\ref{f:PoleFigure_qmap} (right). The
three peaks observed correspond, from top to bottom, to the \LSMO\
electrode, the substrate and the \PTO\ thin film. As can be seen,
the peaks are perfectly aligned in the vertical direction, implying
that the $a$(and $b$)-axis lattice parameters are identical and
equal to the one imposed by the substrate, $a=3.905$\AA . This
demonstrates that the strain state of the \PTO\ films grown on
\LSMO\ is similar to the one of the films grown directly on Nb-doped
\STO\ substrates. Thus, the main difference between the two \PTO\
series is the change of the bottom electrical boundary conditions.

Simulations were then performed to determine the c-axis parameter values, using
$\theta-2\theta$ diffractograms around (00$l$) reflections for $l=1$ to 5
~\cite{lic06_phd}.

\begin{figure}[!ht]
 \begin{center}
  \includegraphics[angle=90,width=\textwidth,keepaspectratio]{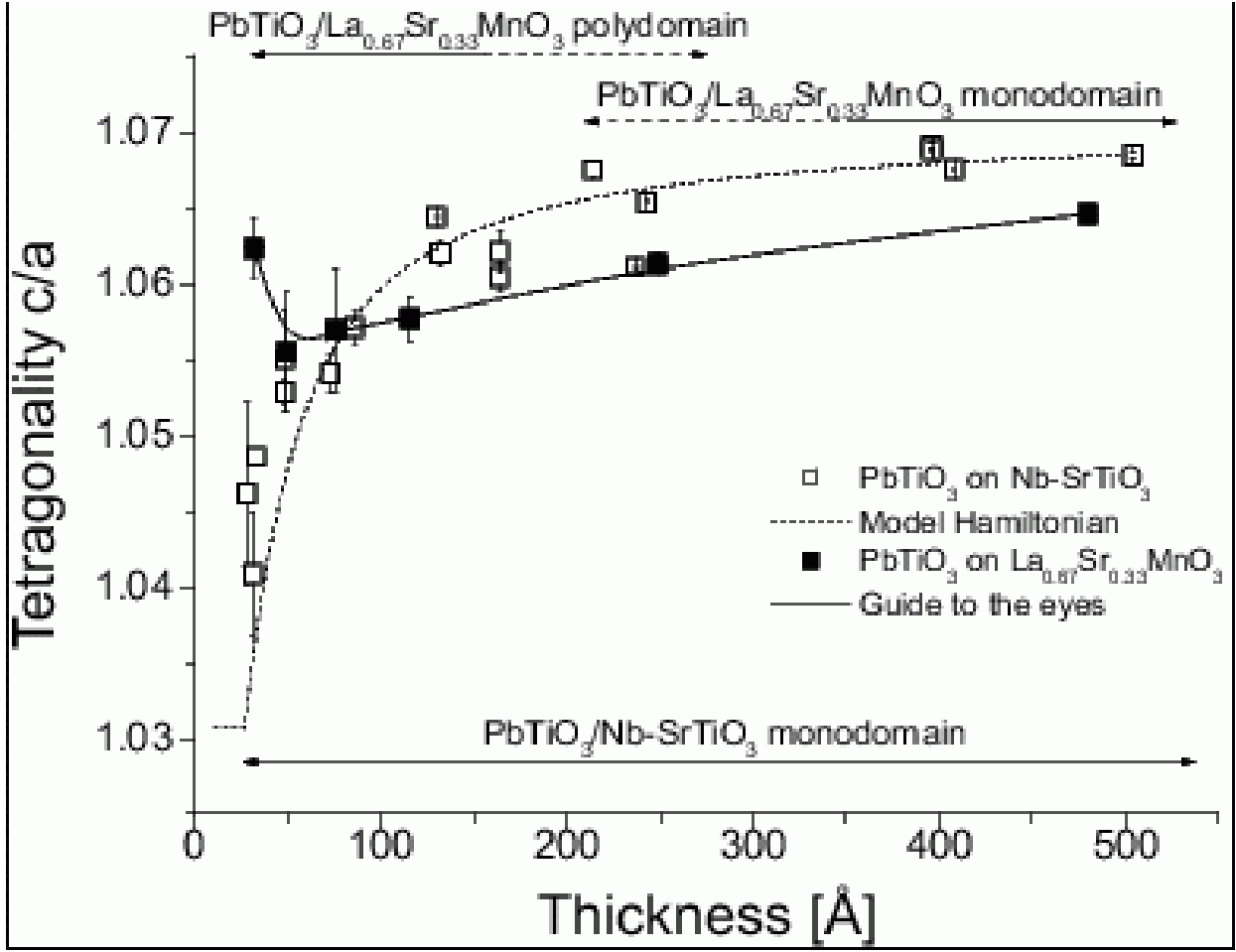}
  \caption{\label{f:ResultsLSMO} Tetragonality as a function of film thickness for  \PTO\ films grown on \LSMO\  (black squares).
  The behavior observed here is different from what was obtained
  in the case of \PTO\ films  grown on Nb-\STO\ (open squares). The model Hamiltonian prediction for monodomain thin
  films with $\lambda_{eff}$=0.12 \AA\
  is shown as a dotted line (details can be found in Ref.~\cite{lic05}).}
\end{center}
\end{figure}

Fig.~\ref{f:ResultsLSMO} shows tetragonality versus thickness for
the two series (the average $c$-axis values and thicknesses obtained
along with their standard deviations are used). The model
Hamiltonian prediction for monodomain thin
  films with $\lambda_{eff}$ = 0.12 \AA\
  is shown as a dotted line~\cite{lic05} and allows the behavior of the series grown on Nb-\STO\ to be explained. As can be seen, the behavior
observed for the thin \PTO\ films grown on \LSMO\ is quite different
from the one observed for \PTO\ films directly grown on Nb-\STO. For
\PTO\ films grown on \LSMO , at large thicknesses where the films
are monodomain (as shown below), the tetragonality measured is
lower than the one of films prepared on Nb-\STO, a possible
indication that the depolarization field is higher (implying a larger
\LSMO\ effective screening length). One would thus expect for this series a substantial decrease of the tetragonality as the film thickness is reduced, with a larger critical thickness. However, as can be seen, the tetragonality is initially only weakly affected by the thickness reduction, with a striking recovery of the
tetragonality for the thinnest film. This non-monotonic behavior
may be a signature of a switching from a ferroelectric monodomain
structure at large thicknesses to a polydomain configuration as the
film thickness is reduced. In a monodomain configuration, when the
film thickness decreases, the depolarization field increases and the
polarization of the monodomain film will decrease (this is what we
observe on Nb-\STO). However, another solution for the system to reduce its
energy is to switch to a polydomain configuration as predicted by
V.~Nagarajan~\etal ~for ultrathin epitaxial
PbZr$_{0.2}$Ti$_{0.8}$O$_3$/\SRO\ heterostructures on \STO\
substrates~\cite{nag06}.

\begin{figure}[!ht]
 \begin{center}
  \includegraphics[angle=90,width=\textwidth,keepaspectratio]{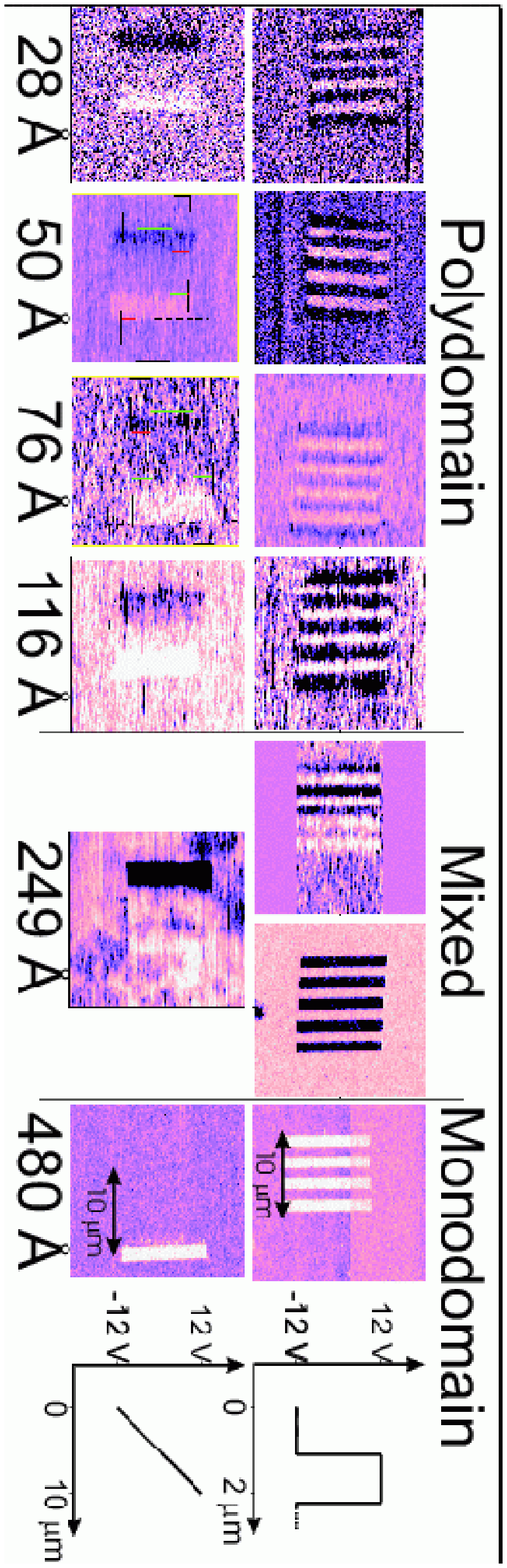}
  \caption{\label{f:AFMpiezo} {\bf Top -} Piezoresponse
  signals obtained after alternate -12V and +12V
  voltages were applied between the metallic tip and the conducting \LSMO\ layer to polarize
  nine stripes over a 10$\times$10 $\mu$m$^2$ area. {\bf Bottom -} Piezoresponse signal obtained after
  application over
  a 10$\times$10 $\mu$m$^2$ square of a voltage gradually ramped from -12V up to +12V. These data also demonstrate ferroelectric
  switching of the polarization in PbTiO$_3$ films as thin as 28 \AA .}
 \end{center}
\end{figure}

 To test this idea
and to probe the domain structure of the different films prepared on
\LSMO , we used the piezoresponse mode of the \afme\
(PFM)~\cite{tyb98, tyb99}. In Fig.~\ref{f:AFMpiezo} are
shown the piezoresponse of the different samples after alternate -12V and +12V voltages were applied between a metallic AFM
tip and the conducting \LSMO\ layer to polarize nine well-defined
stripes over a 10$\times$10$\mu$m$^2$ area (Fig.~\ref{f:AFMpiezo} top) and a
gradual ramp from -12V up to +12V was applied to another area of the
sample (Fig.~\ref{f:AFMpiezo} bottom). By comparing the background signal (unwritten area) to the
one of the written areas, one can deduce whether the sample is mono-
or poly-domain. For the \PTO\ thin films of 28, 50, 76 and 116 \AA ,
the background in Fig.~\ref{f:AFMpiezo} gives a signal corresponding
to the average of the signal given by the areas written with positive
and negative voltages. This strongly suggests that these samples are
polydomain, with domains smaller than the tip resolution (leading to
a signal corresponding to the average over the different up- and
down- domains). In contrast, for the thicker sample (480 \AA ), only
the lines written with a positive voltage applied to the tip can be
seen. This means that the background signal is identical to the
signal given by the lines written with a negative voltage, strongly suggesting that the sample is monodomain with an
up-polarization~\footnote{We note that the polarization direction observed here is opposite to the polarization direction found in the monodomain \PTO\ films grown on Nb-\STO ~\cite{lic05}.} (and thus application of a negative voltage has no
effect). Interestingly, the 249 \AA\ thick sample gives different
images depending on the tip location, one corresponding to a
monodomain background, the other to a polydomain one. In the image
obtained after gradually writing with a ramp from -12V up to +12V,
one sees that, in the background, some large regions appear with
different polarizations. This sample is most probably in a mixed
state with some large areas being monodomain, and others being
polydomain. We note that in contrast to the ``pinned'' polydomain
state of V.~Nagarajan~\etal ~\cite{nag06}, the polydomain state that
we find can be switched with an electric field, and large domains
remain stable for at least 24 hours.

Another measurement that allows the polarization state to be
determined (monodomain up or down, or polydomain) and the
tetragonality of the top few layers to be measured is x-ray
photoelectron diffraction (XPD)~\cite{des06_1,des06_3}, which was
carried out on two samples, 28 \AA\ and 249 \AA\
thick. Curiously (in light of our AFM measurements), during the XPD
measurements, both samples were observed to be monodomain up with a
tetragonality that, in the case of the thinner sample, is much
smaller than the value determined by x-ray diffraction. However,
this low $c/a$ value is found to be in good agreement with the value
obtained using XPD on a monodomain \PTO\ thin film of similar
thickness prepared on a Nb-doped \STO\ substrate. It appears that
during the XPD measurement, the samples are forced to be in a
monodomain configuration, causing a reduction of the polarization
(for very thin films) and a concomitant decrease of the
tetragonality. Once the XPD measurements are finished, the samples
return to their more stable state, which for the thinner film is a
polydomain state, allowing the recovery of the polarization and
therefore also of the tetragonality, as was checked by repeating the
x-ray diffraction and PFM measurements. These experiments thus
demonstrate in thin films the direct relationship between the
tetragonality value and the domain configuration.

To support our experimental observations, we used a simple model
solely for the purpose of identifying key thicknesses at which
changes in domain structure can be expected. The idea behind this
model is that the top surface has a strong preference for one
polarization direction over the other, at the very least during the
growth process, presumably because of the availability of a
particular surface adsorbate. The energy can be written  as $U(P) =
B P^{2} +C P^{4}-\frac{1}{2}E_{dep}\cdot P$ with, according to first
principles calculations (T=0K)~\cite{daw05_1}, $B = - 0.17175279$, $C =
+0.16068441$ (U in eV/unit cell) for PbTiO$_{3}$ constrained in
plane to the lattice parameter of SrTiO$_{3}$. As the film thickness
is decreased, the depolarization field increases, which modifies the
shape of $U(P)$. When looking at a monodomain scenario, one
considers directly the decrease of the polarization to zero as the
depolarization field is increased~\cite{lic05}. A different scenario
is to consider for a fixed polarization state how stable this state
is against the zero polarization state (which locally the system
must pass through to get to a polydomain state)~\footnote{The stability of
the polarization against the opposite polarization state cannot be
assessed directly, as the energetics of adsorbate screening is
difficult to quantify.}. As the thickness decreases and the
depolarization field continues to increase, a point is reached at
which the local minimum in the as-grown state exists but is at the
same or higher energy than the zero polarization state. At this
point, domain walls with zero polarization are energetically more favorable than the as-grown polarization direction, but there is
still an energy barrier, which means that the polarization may be
partially stable. As the thickness is further decreased, eventually
the local minimum disappears and the sample must become polydomain.
The first field is found to be $E_{dep,1}=3.7\times10^8$ V/m with a
polarization of $P_1=60\mu C/cm^2$ and the second is
$E_{dep,2}=5.2\times10^8$ V/m with $P_2=42\mu
C/cm^2$. Assuming perfect screening of the polarization on the top surface by the
adsorbates, the depolarization field is only due to the bottom
electrode and can be written as $E_{dep,i}=\frac{\lambda
P_i}{d_i\epsilon_{0}}$ where $\lambda$ is the electrode effective screening
length and $d_i$ is the film thickness. Using a screening length of
2 {\AA}, as observed for the \LSMO\ thin films (Ref.~\cite{hon05})~\footnote{Note that the effective screening length appearing in the model Hamiltonian, although related, is not directly the Thomas-Fermi screening length, and includes other contributions than the pure electronic screening, as discussed in Ref.~\cite{ger06}.},
these values correspond to critical thicknesses $d_1=370$\AA\ and
$d_2=185$\AA . This predicts that samples above 370 {\AA} can be
monodomain, samples between 370 {\AA} and 185 {\AA} might be
expected to show either monodomain or polydomain behavior, and that
finally any sample below 185 {\AA} must be polydomain, which
corresponds remarkably well with our experimental results in view of
the simplicity of the model. One finally notes that for a substantially smaller effective screening
length (as seems to be the case for the interface between \PTO\ and
Nb-doped \STO ), these critical thicknesses will also be much smaller,
possibly explaining the different behavior observed between the two
series.

Acknowledgements~: We would like to thank Stefano Gariglio for his expertise in x-ray diffraction, and Caroline Mauron and Jill
Guyonnet for their contribution during their summer internship. This work was
supported by the Swiss National Science Foundation through the
National Center of Competence in Research ``Materials with Novel
Electronic Properties-MaNEP'' and Division II, and ESF (Thiox). C.A. acknowledges primary support from the National Science Foundation under Contract No. MRSEC DMR 0520495 and DMR 0134721 and ONR, along with support from the Packard and Sloan Foundations.


\begin{thebibliography}{19}
\expandafter\ifx\csname natexlab\endcsname\relax\def\natexlab#1{#1}\fi
\expandafter\ifx\csname bibnamefont\endcsname\relax
  \def\bibnamefont#1{#1}\fi
\expandafter\ifx\csname bibfnamefont\endcsname\relax
  \def\bibfnamefont#1{#1}\fi
\expandafter\ifx\csname citenamefont\endcsname\relax
  \def\citenamefont#1{#1}\fi
\expandafter\ifx\csname url\endcsname\relax
  \def\url#1{\texttt{#1}}\fi
\expandafter\ifx\csname urlprefix\endcsname\relax\def\urlprefix{URL }\fi
\providecommand{\bibinfo}[2]{#2}
\providecommand{\eprint}[2][]{\url{#2}}

\bibitem[{\citenamefont{Bune et~al.}(1998)\citenamefont{Bune, Fridkin,
  Ducharme, Blinov, Palto, Sorokin, Yudin, and Zlatkin}}]{bun98}
\bibinfo{author}{\bibfnamefont{A.}~\bibnamefont{Bune}},
  \bibinfo{author}{\bibfnamefont{V.}~\bibnamefont{Fridkin}},
  \bibinfo{author}{\bibfnamefont{S.}~\bibnamefont{Ducharme}},
  \bibinfo{author}{\bibfnamefont{L.}~\bibnamefont{Blinov}},
  \bibinfo{author}{\bibfnamefont{S.}~\bibnamefont{Palto}},
  \bibinfo{author}{\bibfnamefont{A.}~\bibnamefont{Sorokin}},
  \bibinfo{author}{\bibfnamefont{S.}~\bibnamefont{Yudin}}, \bibnamefont{and}
  \bibinfo{author}{\bibfnamefont{A.}~\bibnamefont{Zlatkin}},
  \bibinfo{journal}{\nat} \textbf{\bibinfo{volume}{391}}, \bibinfo{pages}{874}
  (\bibinfo{year}{1998}).

\bibitem[{\citenamefont{Tybell et~al.}(1999)\citenamefont{Tybell, Ahn, and
  Triscone}}]{tyb99}
\bibinfo{author}{\bibfnamefont{T.}~\bibnamefont{Tybell}},
  \bibinfo{author}{\bibfnamefont{C.~H.} \bibnamefont{Ahn}}, \bibnamefont{and}
  \bibinfo{author}{\bibfnamefont{J.-M.} \bibnamefont{Triscone}},
  \bibinfo{journal}{\apl} \textbf{\bibinfo{volume}{75}}, \bibinfo{pages}{856}
  (\bibinfo{year}{1999}).

\bibitem[{\citenamefont{Ghosez and Rabe}(2000)}]{gho00}
\bibinfo{author}{\bibfnamefont{P.}~\bibnamefont{Ghosez}} \bibnamefont{and}
  \bibinfo{author}{\bibfnamefont{K.~M.} \bibnamefont{Rabe}},
  \bibinfo{journal}{\apl} \textbf{\bibinfo{volume}{76}}, \bibinfo{pages}{2767}
  (\bibinfo{year}{2000}).

\bibitem[{\citenamefont{Meyer and Vanderbilt}(2001)}]{mey01}
\bibinfo{author}{\bibfnamefont{B.}~\bibnamefont{Meyer}} \bibnamefont{and}
  \bibinfo{author}{\bibfnamefont{D.}~\bibnamefont{Vanderbilt}},
  \bibinfo{journal}{\prb} \textbf{\bibinfo{volume}{63}},
  \bibinfo{pages}{205426} (\bibinfo{year}{2001}).

\bibitem[{\citenamefont{Streiffer et~al.}(2002)\citenamefont{Streiffer,
  Eastman, Fong, Thompson, Munkholm, Murty, Auciello, Bai, and
  Stephenson}}]{str02}
\bibinfo{author}{\bibfnamefont{S.}~\bibnamefont{Streiffer}},
  \bibinfo{author}{\bibfnamefont{J.}~\bibnamefont{Eastman}},
  \bibinfo{author}{\bibfnamefont{D.}~\bibnamefont{Fong}},
  \bibinfo{author}{\bibfnamefont{C.}~\bibnamefont{Thompson}},
  \bibinfo{author}{\bibfnamefont{A.}~\bibnamefont{Munkholm}},
  \bibinfo{author}{\bibfnamefont{M.~R.} \bibnamefont{Murty}},
  \bibinfo{author}{\bibfnamefont{O.}~\bibnamefont{Auciello}},
  \bibinfo{author}{\bibfnamefont{G.}~\bibnamefont{Bai}}, \bibnamefont{and}
  \bibinfo{author}{\bibfnamefont{G.}~\bibnamefont{Stephenson}},
  \bibinfo{journal}{\prl} \textbf{\bibinfo{volume}{89}},
  \bibinfo{pages}{067601} (\bibinfo{year}{2002}).

\bibitem[{\citenamefont{Junquera and Ghosez}(2003)}]{jun03_1}
\bibinfo{author}{\bibfnamefont{J.}~\bibnamefont{Junquera}} \bibnamefont{and}
  \bibinfo{author}{\bibfnamefont{P.}~\bibnamefont{Ghosez}},
  \bibinfo{journal}{\nat} \textbf{\bibinfo{volume}{422}}, \bibinfo{pages}{506}
  (\bibinfo{year}{2003}).

\bibitem[{\citenamefont{Fong et~al.}(2004)\citenamefont{Fong, Stephenson,
  Streiffer, Eastman, Auciello, Fuoss, and Thompson}}]{fon04}
\bibinfo{author}{\bibfnamefont{D.~D.} \bibnamefont{Fong}},
  \bibinfo{author}{\bibfnamefont{G.~B.} \bibnamefont{Stephenson}},
  \bibinfo{author}{\bibfnamefont{S.~K.} \bibnamefont{Streiffer}},
  \bibinfo{author}{\bibfnamefont{J.~A.} \bibnamefont{Eastman}},
  \bibinfo{author}{\bibfnamefont{O.}~\bibnamefont{Auciello}},
  \bibinfo{author}{\bibfnamefont{P.~H.} \bibnamefont{Fuoss}}, \bibnamefont{and}
  \bibinfo{author}{\bibfnamefont{C.}~\bibnamefont{Thompson}},
  \bibinfo{journal}{\sci} \textbf{\bibinfo{volume}{304}}, \bibinfo{pages}{1650}
  (\bibinfo{year}{2004}).

\bibitem[{\citenamefont{Lichtensteiger
  et~al.}(2005)\citenamefont{Lichtensteiger, Triscone, Junquera, and
  Ghosez}}]{lic05}
\bibinfo{author}{\bibfnamefont{C.}~\bibnamefont{Lichtensteiger}},
  \bibinfo{author}{\bibfnamefont{J.-M.} \bibnamefont{Triscone}},
  \bibinfo{author}{\bibfnamefont{J.}~\bibnamefont{Junquera}}, \bibnamefont{and}
  \bibinfo{author}{\bibfnamefont{P.}~\bibnamefont{Ghosez}},
  \bibinfo{journal}{\prl} \textbf{\bibinfo{volume}{94}},
  \bibinfo{pages}{047603} (\bibinfo{year}{2005}).

\bibitem[{\citenamefont{Despont
  et~al.}(2006{\natexlab{a}})\citenamefont{Despont, Lichtensteiger, Koitzsch,
  Clerc, Garnier, {Garcia de Abajo}, Bousquet, Ghosez, Triscone, and
  Aebi}}]{des06_1}
\bibinfo{author}{\bibfnamefont{L.}~\bibnamefont{Despont}},
  \bibinfo{author}{\bibfnamefont{C.}~\bibnamefont{Lichtensteiger}},
  \bibinfo{author}{\bibfnamefont{C.}~\bibnamefont{Koitzsch}},
  \bibinfo{author}{\bibfnamefont{F.}~\bibnamefont{Clerc}},
  \bibinfo{author}{\bibfnamefont{M.~G.} \bibnamefont{Garnier}},
  \bibinfo{author}{\bibfnamefont{F.~J.} \bibnamefont{{Garcia de Abajo}}},
  \bibinfo{author}{\bibfnamefont{E.}~\bibnamefont{Bousquet}},
  \bibinfo{author}{\bibfnamefont{P.}~\bibnamefont{Ghosez}},
  \bibinfo{author}{\bibfnamefont{J.-M.} \bibnamefont{Triscone}},
  \bibnamefont{and} \bibinfo{author}{\bibfnamefont{P.}~\bibnamefont{Aebi}},
  \bibinfo{journal}{\prb} \textbf{\bibinfo{volume}{73}},
  \bibinfo{pages}{094110} (\bibinfo{year}{2006}{\natexlab{a}}).

\bibitem[{\citenamefont{Mehta et~al.}(1973)\citenamefont{Mehta, Silverman, and
  Jacobs}}]{meh73}
\bibinfo{author}{\bibfnamefont{R.~R.} \bibnamefont{Mehta}},
  \bibinfo{author}{\bibfnamefont{B.~D.} \bibnamefont{Silverman}},
  \bibnamefont{and} \bibinfo{author}{\bibfnamefont{J.~T.}
  \bibnamefont{Jacobs}}, \bibinfo{journal}{\jap} \textbf{\bibinfo{volume}{44}},
  \bibinfo{pages}{3379} (\bibinfo{year}{1973}).

\bibitem[{\citenamefont{Batra et~al.}(1973)\citenamefont{Batra, Wurfel, and
  Silverman}}]{bat73_3}
\bibinfo{author}{\bibfnamefont{I.~P.} \bibnamefont{Batra}},
  \bibinfo{author}{\bibfnamefont{P.}~\bibnamefont{Wurfel}}, \bibnamefont{and}
  \bibinfo{author}{\bibfnamefont{B.~D.} \bibnamefont{Silverman}},
  \bibinfo{journal}{\jvst} \textbf{\bibinfo{volume}{10}}, \bibinfo{pages}{687}
  (\bibinfo{year}{1973}).

\bibitem[{\citenamefont{Dawber et~al.}()\citenamefont{Dawber, Stucki,
  Lichtensteiger, and Triscone}}]{daw06}
\bibinfo{author}{\bibfnamefont{M.}~\bibnamefont{Dawber}},
  \bibinfo{author}{\bibfnamefont{N.}~\bibnamefont{Stucki}},
  \bibinfo{author}{\bibfnamefont{C.}~\bibnamefont{Lichtensteiger}},
  \bibnamefont{and} \bibinfo{author}{\bibfnamefont{J.-M.}
  \bibnamefont{Triscone}}, \bibinfo{howpublished}{to be published}.

\bibitem[{\citenamefont{Lichtensteiger}(2006)}]{lic06_phd}
\bibinfo{author}{\bibfnamefont{C.}~\bibnamefont{Lichtensteiger}}, Ph.D. thesis,
  \bibinfo{school}{University of Geneva} (\bibinfo{year}{2006}).

\bibitem[{\citenamefont{Nagarajan et~al.}(2006)\citenamefont{Nagarajan,
  Junquera, He, Jia, Waser, Lee, Kim, Baik, Zhao, Ramesh et~al.}}]{nag06}
\bibinfo{author}{\bibfnamefont{V.}~\bibnamefont{Nagarajan}},
  \bibinfo{author}{\bibfnamefont{J.}~\bibnamefont{Junquera}},
  \bibinfo{author}{\bibfnamefont{J.~Q.} \bibnamefont{He}},
  \bibinfo{author}{\bibfnamefont{C.~L.} \bibnamefont{Jia}},
  \bibinfo{author}{\bibfnamefont{R.}~\bibnamefont{Waser}},
  \bibinfo{author}{\bibfnamefont{K.}~\bibnamefont{Lee}},
  \bibinfo{author}{\bibfnamefont{Y.~K.} \bibnamefont{Kim}},
  \bibinfo{author}{\bibfnamefont{S.}~\bibnamefont{Baik}},
  \bibinfo{author}{\bibfnamefont{T.}~\bibnamefont{Zhao}},
  \bibinfo{author}{\bibfnamefont{R.}~\bibnamefont{Ramesh}},
  \bibnamefont{et~al.}, \bibinfo{journal}{\jap} \textbf{\bibinfo{volume}{100}},
  \bibinfo{pages}{051609} (\bibinfo{year}{2006}).

\bibitem[{\citenamefont{Tybell et~al.}(1998)\citenamefont{Tybell, Ahn, and
  Triscone}}]{tyb98}
\bibinfo{author}{\bibfnamefont{T.}~\bibnamefont{Tybell}},
  \bibinfo{author}{\bibfnamefont{C.~H.} \bibnamefont{Ahn}}, \bibnamefont{and}
  \bibinfo{author}{\bibfnamefont{J.-M.} \bibnamefont{Triscone}},
  \bibinfo{journal}{\apl} \textbf{\bibinfo{volume}{72}}, \bibinfo{pages}{1454}
  (\bibinfo{year}{1998}).

\bibitem[{\citenamefont{Despont
  et~al.}(2006{\natexlab{b}})\citenamefont{Despont, Lichtensteiger, Clerc,
  Garnier, {Garcia de Abajo}, {Van Hove}, Triscone, and Aebi}}]{des06_3}
\bibinfo{author}{\bibfnamefont{L.}~\bibnamefont{Despont}},
  \bibinfo{author}{\bibfnamefont{C.}~\bibnamefont{Lichtensteiger}},
  \bibinfo{author}{\bibfnamefont{F.}~\bibnamefont{Clerc}},
  \bibinfo{author}{\bibfnamefont{M.~G.} \bibnamefont{Garnier}},
  \bibinfo{author}{\bibfnamefont{F.~J.} \bibnamefont{{Garcia de Abajo}}},
  \bibinfo{author}{\bibfnamefont{M.~A.} \bibnamefont{{Van Hove}}},
  \bibinfo{author}{\bibfnamefont{J.-M.} \bibnamefont{Triscone}},
  \bibnamefont{and} \bibinfo{author}{\bibfnamefont{P.}~\bibnamefont{Aebi}},
  \bibinfo{journal}{\epjb} \textbf{\bibinfo{volume}{49}}, \bibinfo{pages}{141}
  (\bibinfo{year}{2006}{\natexlab{b}}).

\bibitem[{\citenamefont{Dawber et~al.}(2005)\citenamefont{Dawber,
  Lichtensteiger, Cantoni, Veithen, Ghosez, Johnston, Rabe, and
  Triscone}}]{daw05_1}
\bibinfo{author}{\bibfnamefont{M.}~\bibnamefont{Dawber}},
  \bibinfo{author}{\bibfnamefont{C.}~\bibnamefont{Lichtensteiger}},
  \bibinfo{author}{\bibfnamefont{M.}~\bibnamefont{Cantoni}},
  \bibinfo{author}{\bibfnamefont{M.}~\bibnamefont{Veithen}},
  \bibinfo{author}{\bibfnamefont{P.}~\bibnamefont{Ghosez}},
  \bibinfo{author}{\bibfnamefont{K.}~\bibnamefont{Johnston}},
  \bibinfo{author}{\bibfnamefont{K.~M.} \bibnamefont{Rabe}}, \bibnamefont{and}
  \bibinfo{author}{\bibfnamefont{J.-M.} \bibnamefont{Triscone}},
  \bibinfo{journal}{Physical Review Letters} \textbf{\bibinfo{volume}{95}},
  \bibinfo{eid}{177601} (\bibinfo{year}{2005}).

\bibitem[{\citenamefont{Hong et~al.}(2005)\citenamefont{Hong, Posadas, and
  Ahn}}]{hon05}
\bibinfo{author}{\bibfnamefont{X.}~\bibnamefont{Hong}},
  \bibinfo{author}{\bibfnamefont{A.}~\bibnamefont{Posadas}}, \bibnamefont{and}
  \bibinfo{author}{\bibfnamefont{C.~H.} \bibnamefont{Ahn}},
  \bibinfo{journal}{Applied Physics Letters} \textbf{\bibinfo{volume}{86}},
  \bibinfo{eid}{142501} (\bibinfo{year}{2005}).

\bibitem[{\citenamefont{Gerra et~al.}(2006)\citenamefont{Gerra, Tagantsev,
  Setter, and Parlinski}}]{ger06}
\bibinfo{author}{\bibfnamefont{G.}~\bibnamefont{Gerra}},
  \bibinfo{author}{\bibfnamefont{A.~K.} \bibnamefont{Tagantsev}},
  \bibinfo{author}{\bibfnamefont{N.}~\bibnamefont{Setter}}, \bibnamefont{and}
  \bibinfo{author}{\bibfnamefont{K.}~\bibnamefont{Parlinski}},
  \bibinfo{journal}{\prl} \textbf{\bibinfo{volume}{96}},
  \bibinfo{pages}{107603} (\bibinfo{year}{2006}).

\end{thebibliography}

\end{document}